\documentclass[a4paper,11pt]{article}
\usepackage{amsmath,amssymb}
\usepackage{slashed}
\usepackage{jheppub}
\usepackage[T1]{fontenc}
\usepackage[utf8]{inputenc}
\usepackage{simplewick}	
\usepackage{graphics}
\usepackage{float}
\usepackage{appendix}
\numberwithin{equation}{section} 
\DeclareMathOperator{\Tr}{Tr}	
\usepackage{setspace}
\newcommand{\be}{\begin{equation}}
\newcommand{\ee}{\end{equation}}
\newcommand{\ben}{\begin{eqnarray}}
\newcommand{\een}{\end{eqnarray}}
\newcommand{\nn}{\nonumber}
\newcommand{\p}{\partial}

\def\l{\left}
\def\r{\right}

\title{$\epsilon$-Expansion in the Gross-Neveu Model from Conformal Field Theory}

\author[a]{Sudip Ghosh,}
\author[b]{Rajesh Kumar Gupta,}
\author[a]{Kasi Jaswin,}
\author[a]{and Amin A. Nizami}

\vspace{5mm}

\affiliation[a] {\it International Centre for Theoretical Sciences, TIFR, Hesaraghatta, Hubli, Bengaluru-560089, India }
\affiliation[b]
{\it International Centre for Theoretical Physics, Strada Costiera 11, 34151 Trieste, Italy}

\vspace{1cm}

\emailAdd{sudip112phys@gmail.com}
\emailAdd{rgupta@ictp.it }
\emailAdd{jaswin@icts.res.in}
\emailAdd{amin@icts.res.in}

\abstract{ We compute the anomalous dimensions of a class of operators of the form $(\bar\psi\psi)^p$ and $(\bar\psi\psi)^p\psi$ to leading order in $\epsilon$ in the Gross-Neveu model in $2+\epsilon$ dimensions. We use the technique developed in arXiv: 1505.00963.}

\preprint{\parbox{3cm}{ICTS/2015/11 }}

\begin{document} 
\maketitle
\flushbottom

\section{Introduction}
In recent work \cite{Rychkov} (see also \cite{Chetan}) the techniques of conformal field theory have been used for the computation of leading order anomalous dimensions  of composite operators in interacting CFTs defined in terms of epsilon expansions about $d=3,4$ spacetime dimensions. The novelty of this technique lies in using conformal symmetry judiciously without taking recourse to any perturbative methods and Feynman diagrams, which has so far been used in such calculations.

The goal of this work is to compute the leading order - in the epsilon expansion - anomalous dimensions of a class of composite operators in the Gross- Neveu model in $2+\epsilon$ dimensions. Our analysis involves two and three point functions and the OPE of relevant operators, and uses only conformal symmetry. We thus accomplish this without relying on Feynman diagrams and conventional perturbation theory techniques. The analysis follows closely the methods of \cite{Rychkov} who first used the method to determine anomalous dimensions of similar operators in the $O(N)$ vector model. 

This note is organised as follows. We provide the basic set up in section \ref{GN}. In section \ref{V1V2} we use methods similar to \cite{Rychkov} to compute the anomalous dimensions of the operators $\psi$ and $\bar{\psi} \psi$. The result of this section is in agreement with that available in the literature. After this simple illustration of the technique, we turn to the general case of higher composite operators. In the appendix we compute the required combinatorial coefficients in the free theory OPE using a recursive diagrammatic approach \cite{Chetan}. In section \ref{V2p}, the two and three point functions, as well as the OPE, of the interacting theory are used and matching with the expected free theory results ultimately leads to a pair of recursion relations involving the leading order anomalous dimensions. The final result for the leading order anomalous dimensions are given in equation (\ref{final}). In section 5, we compute the anomalous dimensions of scalars which are not singlet under $U(\tilde N)$. As far we know, these have not been computed before in the literature and the results of section 4 and 5 are new.

\section{The Gross-Neveu model} \label{GN}
The Gross-Neveu model \cite{GrossNeveu} is a renormalizable field theory in two dimensions. It is described by a $U(\tilde N)$ symmetric action for $\tilde N$ massless self-interacting Dirac fermions $\{\psi^I,\bar\psi^I\}$. We will consider the Gross-Neveu model in $2+\epsilon$ dimensions \cite{Moshe}
\be
S=\int d^{2+\epsilon}x\l[\bar\psi^I\slashed \p\psi^I+\frac{1}{2}g\mu^{-\epsilon}\l(\bar\psi^I\psi^I\r)^2\r],\quad I=1,....,\tilde N.
\ee
Here $g$ is the coupling constant which is dimensionless in two dimensions. This theory has a weakly coupled UV fixed point given by the non-trivial zero of the beta function,
\be
\beta(g)=\epsilon g-\l(N-2\r)\frac{g^2}{2\pi},\qquad N= \tilde N\Tr{\mathbb{I}}\,.
\ee
Here $\Tr{\mathbb{I}}$ is the trace of identity in Dirac fermion space, and in two dimensions $N=2\tilde N$. The fixed point occurs at 
\be
g_*=\frac{2\pi\epsilon}{N-2}+\mathcal O(\epsilon^2)\,.
\ee
The special case of $N=2$ for which the $\beta$ function vanishes identically corresponds to the Thirring model. In this paper we consider the case for which $N>2$.\\
The dimensions of the fermion $\psi^I$, $\Delta_1$, and composite scalar $\bar\psi^I\psi^I$, $\Delta_2$, are given by
\ben
&&\Delta_1=\frac{d-1}{2}+\gamma_1=\frac{1}{2}+\frac{\epsilon}{2}+\gamma_1\,,\nn\\
&&\Delta_2=d-1+\gamma_2=1+\epsilon+\gamma_2\,.
\een
The anomalous dimensions of the fundamental fermions and the composite scalar in the $\epsilon$-expansion have been computed in perturbation theory using the standard Feynman diagram techniques and, to leading order in $\epsilon$, are given by
\ben\label{anomdim1}
&& \gamma_1=\frac{N-1}{16\pi^2}g_*^2=\frac{\l(N-1\r)\epsilon^2}{4\l(N-2\r)^2}\,,\nn\\
&&\gamma_2=-\frac{N-1}{2\pi}g_*=-\frac{N-1}{N-2}\epsilon\,.
\een
The purpose of this note is to derive the above expressions, and similar ones for higher dimensional composite operators,  using conformal field theory techniques without doing Feynman diagram computations\footnote{See \cite{N1, N2} for various aspects of the Gross-Neveu model in $2+\epsilon$ dimensions}. For this we assume that the fixed point is a conformal fixed point. \\ 
In two dimensions, the fermion propagator is given by
\be
\l<\psi^I(x)\bar\psi^J(y)\r>=\frac{\delta^{IJ}}{2\pi} \frac{\Gamma^\mu{(x-y)_\mu}}{(x-y)^{2}}\,.
\ee
We normalise our fields 
\be
\psi^I_{\text {new}}=\sqrt{2\pi}\psi^I,\quad \bar\psi^I_{\text{new}}=\sqrt{2\pi}\bar\psi^I\,.
\ee
In order to simplify the notation in the analysis below, we use the normalised elementary field but denoted by the old variable.  In this normalisation, the two point function is 
\be
\l<\psi^I(x)\bar\psi^J(y)\r>=\delta^{IJ} \frac{\Gamma^\mu{(x-y)_\mu}}{(x-y)^{2}}\,,
\ee
and the equation of motions are\footnote{In general, for non-integer dimensions, gamma matrices are infinite dimensional and there are infinite number of antisymmetrized products. However for the calculation of anomalous dimensions to the leading order in $\epsilon$, for the class of operators $(\bar\psi\psi)^n$ and $\psi(\bar\psi\psi)^n$, this complication will not play any role. }
\ben
&&\slashed\p\psi^I=-\frac{g\mu^{-\epsilon}}{2\pi}\psi^I\l(\bar\psi^J\psi^J\r)\,,\\
&&\p_\mu\bar\psi^I\Gamma^\mu=\frac{g\mu^{-\epsilon}}{2\pi}\bar\psi^I\l(\bar\psi^J\psi^J\r)\,.
\een 
In the free theory the fermions satisfy $\slashed\p\psi^I=0,\, \p_\mu\bar\psi^I\Gamma^\mu=0$ which are the shortening conditions for the multiplets $\{\psi^I\}_{\text {free}}$ and $\{\bar\psi^I\}_{\text{free}}$. In addition all other bilinears of $\psi^I$ and $\bar\psi^I$ are primary operators. At the interacting fixed point $\{\psi^I\}_{\text {fixed pt}}$ and $\{\bar\psi^I\}_{\text {fixed pt}}$ are no longer short multiplets. The primary operators in the free theory $\psi^I\l(\bar\psi^J\psi^J\r)$ and $\bar\psi^I\l(\bar\psi^J\psi^J\r)$ now become descendants of the $\{\psi^I\}_{\text {fixed pt.}}$ and $\{\bar\psi^I\}_{\text {fixed pt.}}$ respectively.
This phenomena of multiplet recombination was observed in $\phi^4$-theory \cite{Rychkov} where two conformal multiplets in the free theory join and become a single conformal multiplet at the interacting fixed point.\\
As in \cite{Rychkov} we assume that every operator $\mathcal O$ in the free theory has a counterpart $V_{\mathcal O}$ at the interacting fixed point. The operators $V_{\mathcal O}$ and their correlation functions in the interacting theory, approach, respectively, $\mathcal O$ and their free correlation function in the $\epsilon\rightarrow 0$ limit. In the Gross Neveu model at the IR free point, various operators are constructed out of products of elementary operators $\psi$ and $\bar\psi$. We will denote operators in the interacting theory as $V_{2p}$, $V_{2p+1}$ and $\bar V_{2p+1}$ such that in the limit $\epsilon \rightarrow 0$ (IR free point)
\begin{equation}
V_{2p}\rightarrow (\bar{\psi}\psi)^{p}, \hspace{0.4cm} V^I_{2p+1}\rightarrow (\bar{\psi}\psi)^{p}\psi^I\,, \hspace{0.4cm} \bar V^I_{2p+1}\rightarrow (\bar{\psi}\psi)^{p}\bar\psi^I\,.
\end{equation} 
We also require that the multiplet recombination is achieved by
\ben\label{eqnprimary}
\slashed\p V_1^I=\alpha V_3^I,\quad \p_\mu \bar{V}_1^I\Gamma^\mu=-\alpha\bar{V}_3^I\,,
\een
for some unknown function $\alpha\equiv \alpha(\epsilon)$ which will be determined below. As an equation of motion, this follows from the Gross-Neveu lagrangian, but in the non-lagrangian approach we follow it is to be interpreted purely as an operator relation indicating that the operator $V_3$ is, in the interacting theory, a descendant of the primary operator $V_1$.

Let us illustrate, schematically, how multiplet recombination is used together with the OPE to determine the leading order anomalous dimensions (This is the method developed by \cite{Rychkov} and used in later sections here).  Suppose the interacting theory has an operator relation of the form $\partial V_{O}=\alpha V_{O'}$ (as explained above, $O$'s denote operators in the free theory whose counterparts in the interacting theory are the $V_{O}$ 's ). We first find an OPE in the free theory of primary composite operators $O_{p},O_{p'}$  which contains,  in the leading terms, $O$ and $O'$ :
\begin{equation}
O_{p}(x)\times O_{p'}(0) \supset (O+.....)+\rho (O'+.....)+ subleading\,\, terms.
\end{equation} 
The dots above denote descendant terms (derivatives acing on $O$) and other subleading terms contain other operators in the spectrum (we have supressed various powers of $x$). The leading order terms suffice for our purpose. $\rho$ is a combinatorial coefficient, determined by Wick contractions in the free theory (see the appendix). Now in the interacting theory, the corresponding OPE would read: 
\begin{equation}
V_{O_{p}}(x)\times V_{O_{p'}}(0) \supset (V_{O}+q\, \partial V_{O} .....)+ subleading\,\, terms.
\end{equation} 
Note that, using the operator relation, the second term on the right hand side above is proportonal to $V_{O'}$. In the interacting theory, this operator ($V_{O'}$) is a descendant and this crucial fact, together with matching with the free theory in the $\epsilon \rightarrow 0$ limit, implies 
$q\alpha=\rho $. The coefficient $q$ will be determined later in terms of anomalous dimensions of the operators in the OPE and will be seen to be singular in the $\epsilon \rightarrow 0$ limit. The coefficient $\alpha$ is determined below (to the required leading order in $\epsilon$) and the above relation will be seen (in section 4.3) to lead to a recursion relation for the leading order anomalous dimensions. 

We turn now to the determination of $\alpha$. We have
\be
\l<V^I_1(x)\bar V^J_1(y)\r>=\delta^{IJ}\frac{\Gamma^\mu(x-y)_\mu}{(x-y)^{2\Delta_1+1}}\,.
\ee
Differentiating the above expression and contracting with $\Gamma^\mu$ matrices, we get
\be\label{equforh1}
\l<\slashed\p V^I_1(x)\p_\sigma \bar V^J_1(y)\Gamma^\sigma\r>=\delta^{IJ}\frac{\Gamma^\mu(x-y)_\mu}{(x-y)^{2\Delta_1+3}}h\,,
\ee
where, using $ \Delta_1=\frac{d-1}{2}+\gamma_1$, we get
\be
h=(2\Delta_1+1)({2-d+2d-2\Delta_1-3})\sim -4\gamma_1\,.
\ee
Now requiring that in the limit $\epsilon\rightarrow 0$, $\l<V^I_3(x)\bar V^J_3(y)\r>$ approaches the free theory correlation function
\be\label{}
\l<\psi^I(\bar\psi^K\psi^K)(x)\bar\psi^J(\bar\psi^L\psi^L)(y)\r>=\delta^{IJ}\frac{\Gamma^\mu(x-y)_\mu (N-1)}{(x-y)^4}\,,
\ee
we get the expression for $\alpha$
\be
\alpha=2\sigma\l(\frac{\gamma_1}{N-1}\r)^{1/2},\qquad \sigma=\pm 1\,.
\ee

\section{Anomalous dimension of $\psi$, $\bar\psi$ and $\bar\psi\psi$}\label{V1V2}
In this section we will compute the anomalous dimensions of the fundamental fermion and the composite scalar. The results of this section are in perfect agreement with the leading order anomalous dimension computed from Feynman diagram techniques. In the next section we will generalise this to higher dimensional operators and derive some new results for the anomalous dimensions. \\
We consider the OPE between $\psi$ and $\bar\psi\psi$ in the free theory. \footnote{The OPE coefficients are determined in the free theory using Wick's contraction- see the appendix.} For this we do not need the full OPE except those terms which are sensitive to the multiplet recombination,
\be\label{freeOPE}
\psi^I(x)\times (\bar\psi^J\psi^J)(0)\supset \frac{1}{x^2}\{\slashed{x}\psi^I(0)+ x^2\psi^I(\bar\psi^J\psi^J)(0)+..\}\,.
\ee
We will compare the above expression for the free OPE with the OPE at the interacting UV fixed point. For this we need the three point function at the interacting fixed point. According to \cite{Petkou}, we have\footnote{As we will explain in the next section, in general conformal invariance requires the presence of another term in the 3-point function. However this extra term does not contribute in the calculation of this section.}
\be\label{3ptfunc}
\l<V^I_1(x_1)\bar V^K_1(x_2)V_2(x_3)\r>=\frac{f\, \slashed{x}_{13}\slashed{x}_{23}\delta^{IK}}{\l(x_{12}^2\r)^{\Delta_1-\frac{1}{2}\Delta_2}\l(x_{13}^2x_{23}^2\r)^{\frac{1}{2}\Delta_2+\frac{1}{2}}}\,.
\ee
In the above $f$ is a constant. From this we can compute the following OPE 
\be\label{fixedptOPE}
V^I_1(x_1)\times V_2(x_3)\supset \frac{f\slashed{x}_{13}}{(x_{13}^2)^{\frac{1}{2}\Delta_2+\frac{1}{2}}}C(x_{13},\p_z)V^I_1(z)|_{z=x_3}+.....
\ee
Here 
\be
C(x_{13},\p_z)=A+\l(B_1 x_{13}^\mu+B_2\slashed{x}_{13}\Gamma^\mu\r)\p_\mu+\l(C_1 x_{13}^\mu x_{13}^\nu+C_2x_{13}^\mu\slashed{x}_{13}\Gamma^\nu+C_3x_{13}^2\Gamma^\mu\Gamma^\nu\r)\p_\mu\p_\nu+.........
\ee
It is important to note here that in the above OPE, we have kept only contributions coming from the conformal family of $V^I_1$ which includes $V_3$ as its descendant. Although the multiplet recombination does not necessarily require a lagrangian description, here our knowledge of the shortening condition follows from the equation of motion (\ref{eqnprimary}). Even though we use different methods, we are dealing with perturbative fixed points which do have a lagrangian description. Thus our analysis is not entirely lagrangian independent\footnote{Presumably the shortening condition can be used for more general fixed points to fix values of anomalous dimensions. It would be very interesting to understand the phenomena of multiplet recombination more generally.}.\\
$A,B_{i},C_{i},..$ are functions of conformal dimensions which we determine by considering $x_1\rightarrow x_3$ and expanding (\ref{3ptfunc}) in powers of $x_{13}$,
\ben
\frac{f\, \slashed{x}_{13}\slashed{x}_{23}\delta^{IK}}{\l(x_{12}^2\r)^{\Delta_1-\frac{1}{2}\Delta_2}\l(x_{13}^2x_{23}^2\r)^{\frac{1}{2}\Delta_2+\frac{1}{2}}}&=&\frac{f\, \slashed{x}_{13}\slashed{x}_{23}\delta^{IK}}{(x_{13}^2)^{\frac{1}{2}\Delta_2+\frac{1}{2}}(x_{23}^2)^{\Delta_1+\frac{1}{2}}}\Big[1+2\l(\Delta_1-\frac{1}{2}\Delta_2\r)\frac{x_{23}\cdot x_{13}}{x_{23}^2}\nn\\&&-\l(\Delta_1-\frac{1}{2}\Delta_2\r)\frac{x_{13}^2}{x_{23}^2}+\frac{1}{2}(2\Delta_1-\Delta_2)(2+2\Delta_1-\Delta_2)\frac{(x_{23}\cdot x_{13})^2}{(x_{23}^2)^2}+...\Big]\,.\nn\\
\een
Comparing with (\ref{fixedptOPE}) we can get all the coefficients. We list here the first few coefficients 
\ben\label{coeffconfdimension}
&&A=-1,\quad B_1=-\frac{ \l(\Delta_1-\frac{1}{2}\Delta_2\r)}{\Delta_1+\frac{1}{2}},\quad B_2=\frac{B_1}{2\Delta_1+1-d},\nn\\
&&C_1=-\frac{\l(2\Delta_1-\Delta_2\r)\l(2+2\Delta_1-\Delta_2\r)}{2\l(2\Delta_1+1\r)\l(2\Delta_1+3\r)}\,,\nn\\
&&C_2=\frac{2C_1}{2\Delta_1+1-d},\quad C_3=\frac{1}{2\Delta_1+1-d}\l[C_1+\frac{\l(\Delta_1-\frac{1}{2}\Delta_2\r)}{2\Delta_1+1}\r]\,.
\een
Now we consider the following free correlators in the limit $|x_{1}|<<|x_{2}|$,
\ben
&&\l<\psi^I(x_{1})(\bar\psi^J\psi^J)(0)\bar\psi^K(x_{2})\r>\sim \frac{\slashed{x}_{1}}{x^2_{1}}\l<\psi^I(0)\bar\psi^K(x_{2})\r>\,,\nn\\
&&\l<\psi^I(x_{1})(\bar\psi^J\psi^J)(0)\bar\psi^K(\bar\psi^L\psi^L)(x_{2})\r>\sim\l<\psi^I(\bar\psi^J\psi^J)(0)\bar\psi^K(\bar\psi^L\psi^L)(x_{2})\r>\,.
\een
Using the OPE (\ref{fixedptOPE}), we have 
\ben
\l<V^I_1(x_{1})V_2(0)\bar V^K_1(x_{2})\r>\sim \frac{A f\slashed{x}_{1}}{(x^2_{1})^{\frac{1}{2}\Delta_2+\frac{1}{2}}}\l<V^I_1(0)\bar V^K_1(x_{2})\r>\,.
\een
This will match with the free correlator if $f\rightarrow -1$ in the limit $\epsilon\rightarrow 0$. 
Next we compare the correlation function with the insertion of the descendant operator $\bar{V}_3^I$,
\be\label{corrldecendent1}
\l<V^I_1(x_{1})V_2(0)\bar V_3^I(x_{2})\r>\sim \frac{f\slashed{x}_{1}}{(x^2_{1})^{\frac{1}{2}\Delta_2+\frac{1}{2}}}C(x_{1},\p_\mu)\l<V^I_1(0)\bar V^K_3(x_{2})\r>\,.
\ee
Here $\bar V_3^K$ is the descendant of $\bar V_1^K$ defined in (\ref{eqnprimary}) and the derivative acts on the first insertion. It is very easy to see that the first two terms containing $A,B_1$ in the expansion of $C$ on the right hand side go to zero as we take $\epsilon\rightarrow 0$,
\be
\l<V^I_1(0)\bar V^K_3(x_{2})\r>=-\frac{1}{\alpha}\l<V^I_1(0)\p_\mu \bar V^K_3(x_{2})\r>\Gamma^\mu=-\frac{\delta^{IK}}{(z^2)^{\Delta_1+\frac{1}{2}}}\frac{\sqrt{\gamma_1(N-1)}}{2\pi\sigma}\,.
\ee
Now we see that the contribution to (\ref{corrldecendent1}) will come from the term with $B_2$. In fact using the expansion we get
\ben
\l<V^I_1(x_{1})V_2(0)\bar V^K_3(x_{2})\r>&\sim&\frac{f\slashed{x}_{1}}{(x^2_{1})^{\frac{1}{2}\Delta_2+\frac{1}{2}}} B_2(\slashed{x}_{1}\,\Gamma^\mu\p_\mu)\l<V^I_1(0)\bar V^K_3(x_{2})\r>\,,\nn\\&&=\frac{f B_2\alpha}{(x^2_{1})^{\frac{1}{2}\Delta_2-\frac{1}{2}}}\l<V^I_3(0)\bar V^K_3(x_{2})\r>\,.
\een
In the above we used the equation of motion for the primary field  (\ref{eqnprimary}). Thus we see that it will go to the free correlator if $f B_2\alpha\sim \mathcal O(1)$ in the limit $\epsilon\rightarrow 0$. Since $f$ goes to constant and $\alpha$ goes to zero, $B_2$ must diverge. We also see from (\ref{coeffconfdimension}) that $B_2$ has a chance of blowing up. If we define
\be\label{relationconfdim2}
\delta=\frac{d-1}{2},\quad \Delta_1=\delta+\gamma_1,\quad \Delta_2=2\delta+\gamma_2\,,
\ee
then
\be
B_2\sim -\frac{(\gamma_1-\frac{1}{2}\gamma_2)}{2\gamma_1(\delta+\gamma_1+\frac{1}{2})}\,.
\ee
Thus $B_2$ will blow up if $\gamma_1$ vanishes as at least $\mathcal O(\epsilon^2)$\\
Now we write
\be\label{relationconfdim3}
\gamma_1\sim y_{1,2}\epsilon^2,\quad \gamma_2\sim y_{2,1}\epsilon\,.
\ee
Then we get
\be
fB_2\alpha\sim \frac{y_{2,1}\sigma \, f}{2\sqrt{y_{1,2}(N-1)}}\rightarrow 1\,.
\ee
Using that $f\rightarrow -1$, we get
\be\label{relationY1}
y_{2,1}=-2\sigma\sqrt{y_{1,2}(N-1)}\,.
\ee
Also in the interacting theory, the conformal dimension $\Delta_3$ of the descendant $ V_3^I (x_1)$ is related to $\Delta_1$ of $V_1^I(x_1)$ by
\ben
&&\Delta_3=\Delta_1+1\Rightarrow 3\delta+\gamma_3=\delta+\gamma_1+1,\nn\\
&&\gamma_3=\gamma_1-\epsilon\Rightarrow y_{3,1}=-1\,.
\een
We will show this by explicit computation in the next section.\\
Now we are interested in finding the OPE between $ V_3^I$ and $V_2$. This can be obtained from (\ref{fixedptOPE}) by acting with a derivative and using (\ref{eqnprimary}).
\be
V^I_3(x_1)\times V_2(x_3)\supset \frac{\tilde f}{\alpha(x_{13}^2)^{\frac{1}{2}\Delta_2+\frac{1}{2}}}\tilde C(x_{13},\p_z)V^I_1(z)|_{z=x_3}+....
\ee
Here
\be
\tilde C(x_{13},\p_z)=\tilde A+\l(\tilde B_1 x_{13}^\mu+\tilde B_2\slashed{x}_{13}\Gamma^\mu\r)\p_\mu+\l(\tilde C_1 x_{13}^\mu x_{13}^\nu+\tilde C_2x_{13}^\mu\slashed{x}_{13}\Gamma^\nu+\tilde C_3x_{13}^2\Gamma^\mu\Gamma^\nu\r)\p_\mu\p_\nu+.........
\ee
where
\ben
&&\tilde f=(d-\Delta_2-1)f,\quad \tilde A=A,\quad \tilde B_1=\frac{d-\Delta_2+1}{d-\Delta_2-1}B_1,\nn\\ &&\tilde B_2=\frac{1-\Delta_2}{d-\Delta_2-1}B_2-\frac{B_1}{d-\Delta_2-1}\,.
\een
In order to compare with the free correlator, we also need the following OPE
\be
\psi_i(\bar\psi_k\psi_k)(x_{1})\times (\bar\psi_j\psi_j)(0)\supset \frac{1}{ x^2_{1}}\{( N-1)\psi_i(0)+\slashed{x}_{1}\psi_i(\bar\psi_j\psi_j)(0)+..\}
\ee
\ben
&&\l<\psi_i(\bar\psi_k\psi_k)(x_{1})(\bar\psi_j\psi_j)(0)\bar\psi_l(x_{2})\r>\sim \frac{(N-1)}{x^2_{1}}\l<\psi_i(0)\bar\psi_l(x_{2})\r>\,,\nn\\
&&\l<\psi_i(\bar\psi_k\psi_k)(x_{1})(\bar\psi_j\psi_j)(0)\bar\psi_l(\bar\psi_l\psi_l)(x_{2})\r>\sim \frac{\slashed{x}_{1}}{x^2_{1}}\l<\psi_i(\bar\psi_j\psi_j)(0)\bar\psi_l(\bar\psi_l\psi_l)(x_{2})\r>\,.
\een
Proceeding as before, we find that for $|x_{1}|<<|x_{2}|$, we have
\be
\l<V^I_3(x_{1})V_2(0)\bar V^K_3(x_{2})\r>\sim \frac{\tilde f\tilde B_2\slashed{x}_{1}}{(x^2_{1})^{\frac{1}{2}\Delta_2+\frac{1}{2}}}\l< V^I_3(0)\bar V^K_3(x_{2})\r>\,.
\ee
Thus in order to match with the free correlator, we require $\tilde f\tilde B_2\rightarrow 1$. Now using that $f\rightarrow -1$, we get
\be
(1-\Delta_2)B_2-B_1=-1\,.
\ee
Using (\ref{relationconfdim2}) and (\ref{relationconfdim3}), to leading order in $\epsilon$, we get
\be
y_{2,1}+y_{2,1}^2=4y_{1,2}\,.
\ee
Using further (\ref{relationY1}), we get
\be
2y_{1,2}(N-2)=\sigma \sqrt{y_{1,2}( N-1)}\,, 
\ee
which implies
\be
\sigma=+1,\quad y_{1,2}=\frac{(N-1)}{4(N-2)^2},\quad y_{2,1}=-\frac{N-1}{N-2}\,.
\ee
Therefore the anomalous dimensions are
\be
\gamma_1=\frac{( N-1)}{4(N-2)^2}\epsilon^2,\quad \gamma_2=-\frac{N-1}{N-2}\epsilon\,.
\ee
These results are in agreement with results in \cite{Moshe,Bondi}.

\section{Anomalous dimensions of $(\bar{\psi}\psi)^{p}$ and $(\bar{\psi}\psi)^{p}\psi$}\label{V2p}
In this section we will compute the leading order anomalous dimensions of a class of higher dimensional composite operators in the interacting theory described by the UV fixed point of the Gross-Neveu Model. In the free theory limit ($\epsilon \rightarrow 0$) these operators are of the form $(\bar\psi\psi)^p$ and $\psi(\bar\psi\psi)^p$ with $p>1$. Let us denote these operators in the interacting theory as $V_{2p}$ and $V_{2p+1}$ such that in the limit $\epsilon \rightarrow 0$ (axiom) 
\begin{equation}
V_{2p}\rightarrow (\bar{\psi}\psi)^{p}, \hspace{0.4cm} V^I_{2p+1}\rightarrow (\bar{\psi}\psi)^{p}\psi^I\,.
\end{equation}

\subsection{The structure of the OPEs}
We will need the following OPEs in the free theory 
\begin{equation} \label{fope1}
\left(\bar{\psi}\psi\right)^{p}(x_{1}) \times \left(\bar{\psi}\psi\right)^{p}\psi^{I}(0) \supset \frac{f_{2p}}{(x_{1}^{2})^{p}}\left\{\psi^{I}(0)+ \slashed{x}_{1}\rho_{2p}\left(\bar{\psi}\psi\right)\psi^{I}(0)\right\}\,,
\end{equation}
\begin{equation}\label{fope2}
\left(\bar{\psi}\psi\right)^{p}\psi^{I}(x_{1}) \times \left(\bar{\psi}\psi\right)^{p}\left(\bar{\psi}\psi\right)(0) \supset \frac{f_{2p+1}}{(x_{1}^{2})^{p+1}}\left\{\slashed{x}_{1}\psi^{I}(0)+ x_{1}^{2}\rho_{2p+1}\left(\bar{\psi}\psi\right)\psi^{I}(0)\right\}\,,
\end{equation}

where $I$ is an $U(\tilde{N})$ index. $f_{2p}, f_{2p+1}$ and $\rho_{2p}, \rho_{2p+1}$ are combinatorial coefficients. Counting all possible Wick contractions gives their values to be 
\begin{eqnarray}
f_{2p}= \prod_{i=1}^{p} i(N-i), \hspace{0.4cm} f_{2p+1}= (p+1)\prod_{i=1}^{p}i(N-i)\,, \\
\rho_{2p}=-\frac{p}{N-1}, \hspace{0.4cm} \rho_{2p+1}=\frac{N-p-1}{N-1}\,.
\end{eqnarray}
See the appendix for details of the calculation. 
Now let us consider the corresponding OPEs in the interacting theory. The most general structure of the OPE, in the first case where the free theory limit is eq. (\ref{fope1}), is 
\begin{equation} \label{iope1}
V_{2p}(x_{1})\times V_{2p+1}^{I}(0)\supset \left(\frac{1}{(x_{12}^{2})^{a}}C(x_{12},\partial_{2})V_{1}^{I}(x_{2})+\frac{\slashed{x}_{12}}{(x_{12}^{2})^b}D(x_{12},\partial_{2})V_{1}^{I}(x_{2})+...\right)_{x_{2}=0}\,.
\end{equation}
The dots indicate other primary operators that can appear in the OPE. Here 
\begin{eqnarray}
a=\left(\Delta_{2p}+\Delta_{2p+1}-\Delta_{1}\right)/2\,,\\
b=\left(\Delta_{2p}+\Delta_{2p+1}-\Delta_{1}+1\right)/2\,.
\end{eqnarray}
The differential operators $C(x_{12},\partial_2)$ and $D(x_{12},\partial_2)$ have the general form 
\begin{eqnarray}\label{dop}
C(x_{12},\partial_2)= A_{0}+B_{0}x_{12}^{\mu}\partial_{2\mu}+B_{1}\slashed{x}_{12}\slashed{\partial}_{2\mu}+ ....\\
D(x_{12},\partial_2)= A'_{0}+B'_{0}x_{12}^{\mu}\partial_{2\mu}+B'_{1}\slashed{x}_{12}\slashed{\partial}_{2\mu}+ ....
\end{eqnarray}

For the OPE of two generic primary operators (one bosonic and the other fermionic) both of these structures can occur. However now we will show that, for the $V_{2p}(x_1)V_{2p+1}^I(0)$ OPE, only the first structure in eq. (\ref{iope1}) is consistent with our axiomatic requirement that 
in the limit $\epsilon \rightarrow 0$ correlators of the interacting theory should match with corresponding correlators in the free theory. \\

For this consider the 3 pt. function $\left<V_{2p}(x_{1})V_{2p+1}^{I}(x_{2})\bar{V}_{3}^{J}(x_{3})\right>$. Then using the OPE - eq. (\ref{iope1}) - we have,
\begin{eqnarray}
\left<V_{2p}(x_{1})V_{2p+1}^{I}(0)\bar{V}_{3}^{J}(x_{3})\right>_{|x_{1}|\ll|x_{3}|} \sim \left\{\frac{1}{(x_{12}^{2})^a}\left(A_{0}+B_{0}x_{12}^{\mu}\partial_{2\mu}\right)\left<V_{1}^{I}(x_{2})\bar{V}_{3}^{J}(x_{3})\right>+....\right\}_{x_{2}=0}\nn\\
+\left\{\frac{\slashed{x}_{12}}{(x_{12}^{2})^b}\left(A'_{0}+B'_{0}x_{12}^{\mu}\partial_{2\mu}\right)\left<V^{I}_{1}(x_{2})\bar{V}_{3}^{J}(x_{3})\right>+....\right\}_{x_{2}=0}\nn\\
+ \alpha\left\{\frac{B_{1}\slashed{x}_{12}}{(x_{12}^{2})^a}\left<V_{3}^{I}(x_{2})\bar{V}_{3}^{J}(x_{3})\right>+ \frac{B'_{1}x_{12}^{2}}{(x_{12}^{2})^b}\left<V_{3}^{I}(x_{2})\bar{V}_{3}^{J}(x_{3})\right>\right\}_{x_{2}=0}
\end{eqnarray}
 
In the free theory the OPE, eq. (\ref{fope1}) gives,
\begin{equation}
\left<\left(\bar{\psi}\psi\right)^{p}(x_{1})\left(\bar{\psi}\psi\right)^{p}\psi^{I}(0)\left(\bar{\psi}\psi\right)\bar{\psi}^{J}(x_{3})\right>\underset{{|x_{1}|\ll|x_{3}|}}{\sim} \left(\frac{\slashed{x}_{12}}{(x_{12}^{2})^{p}}f_{2p}\rho_{2p}\left<\left(\bar{\psi}\psi\right)\psi^{I}(x_{2})\left(\bar{\psi}\psi\right)\bar{\psi}^{J}(x_{3})\right>\right)_{x_{2}=0}\nonumber
\end{equation}
Now since in the $\epsilon\rightarrow 0$ limit we require 
\begin{equation}
\left<V_{2p}(x_{1})V_{2p+1}^{I}(x_{2})\bar{V}_{3}^{J}(x_{3})\right>\rightarrow \left<\left(\bar{\psi}\psi\right)^{p}(x_{1})\left(\bar{\psi}\psi\right)^{p}\psi^{I}(x_{2})\left(\bar{\psi}\psi\right)\bar{\psi}^{J}(x_{3})\right>\,.
\end{equation}
and, 
\begin{equation}
\left<V_{3}^{I}(x_{2})\bar{V}_{3}^{J}(x_{3}\right>\rightarrow \left<\left(\bar{\psi}\psi\right)\psi^{I}(x_{2})\left(\bar{\psi}\psi\right)\bar{\psi}^{J}(x_{3})\right>\,.
\end{equation}
Hence we clearly see that only the first structure in eqn. (\ref{iope1}) needs to be considered. In other words all the coefficients appearing in $D(x_{12},\partial_{2})$ can be set to zero in this case. \\

Next consider the OPE of $V_{2p+1}^{I}$ and $V_{2p+2}$. Again just on grounds of conformal symmetry we can write down an expression similar to eq. (\ref{iope1}). But once again it is easy to show using the free theory OPE, eq. (\ref{fope2}) that our axiom 
\begin{equation}
\left<V_{2p+1}^{I}(x_{1})V_{2p+2}(x_{2})\bar{V}_{3}^{J}(x_{3})\right> \rightarrow \left<\left(\bar{\psi}\psi\right)^{p}\psi^{I}(x_{1})\left(\bar{\psi}\psi\right)^{p}(\bar{\psi}\psi)(x_{2})\left(\bar{\psi}\psi\right)\bar{\psi}^{J}(x_{3})\right>\,.
\end{equation}
allows only the second structure of eq. (\ref{iope1}) for the OPE of  $V_{2p+1}^{I}$ and $V_{2p+2}$.   \\
Note that the above distinction is important when both operators involved in the OPE are primary operators. When one of the operators is a descendant the structure of the OPE simply follows by acting with derivatives on the OPE of the primary operators. For example when $p=1$ the OPE of $V_{2}$ and $V_{3}^{I}$ can be obtained from the OPE of $V_{1}^{I}$ and $V_{2}$ by differentiating the latter.


\subsection{Determining the coefficients in the OPE}
We will now obtain the expression for the coefficients in eq. (\ref{dop}). The method for doing this is simple. The form of the 3 pt. function which is fixed in the usual way by conformal invariance is matched against the form obtained by taking the OPE of the first two operators within the 3-pt. function. 
We start with the following 3 pt. function, 
\begin{equation}
\left<V_{2p}(x_{1})V_{2p+1}^{I}(x_{2})\bar{V}_{1}^{J}(x_{3})\right>= g_{1} \frac{\slashed{x}_{12}\slashed{x}_{13}\delta^{IJ}}{(x_{12}^{2})^{a+1/2}(x_{23}^{2})^{b}(x_{31}^{2})^{c+1/2}}+ g_{2}\frac{\slashed{x}_{23}\delta^{IJ}}{(x_{12}^{2})^{a}(x_{23}^{2})^{b+1/2}(x_{31}^{2})^{c}}\,.
\end{equation}
where 
\begin{eqnarray}\label{3pt1}
&& a= \frac{\left(\Delta_{2p}+\Delta_{2p+1}-\Delta_{1}\right)}{2},\quad b=\frac{\left(\Delta_{2p+1}+\Delta_{1}-\Delta_{2p}\right)}{2}\,,\\
&&c=\frac{\left(\Delta_{1}+\Delta_{2p}-\Delta_{2p+1}\right)}{2}\,. \nonumber
\end{eqnarray}
The form is determined by conformal invariance which allows for both the above structures\footnote{See, for example, \cite{Fradkin}. Contrast this with Petkou's result \cite{Petkou} where only the first term appears.}.
Now using the OPE 
\begin{equation}
V_{2p}(x_{1})\times V_{2p+1}^{I}(0)\supset \left(\frac{1}{(x_{12}^{2})^{a}}C(x_{12},\partial_{2})V_{1}^{I}(x_{2})\right)_{x_{2}=0}\,.
\end{equation}
we get, 
\ben\label{ope1series}
\left<V_{2p}(x_{1})V_{2p+1}^{I}(0)\bar{V}_{1}^{J}(x_{3})\right>_{|x_{1}|\ll|x_{3}|} &\sim& \left(\frac{1}{(x_{12}^{2})^{a}}C(x_{12},\partial_{2})\left<V_{1}^{I}(x_{2})\bar{V}^{J}_{1}(x_{3})\right>\right)_{x_{2}=0}\nonumber\\
&=&\frac{A_{0}(-\slashed{x}_{3})\delta^{IJ}}{(x_{1}^{2})^{a}(x_{3}^{2})^{\Delta_{1}+1/2}}+\frac{B_{0}\delta^{IJ}}{(x_{1}^{2})^{a}(x_{3}^{2})^{\Delta_{1}+1/2}}\Big(\left(\frac{1}{2}-\Delta_{1}\right)x_{3}^{2}\slashed{x}_{1}\nn\\&&-\left(\frac{1}{2}+\Delta_{1}\right)\slashed{x}_{3}\slashed{x}_{1}\slashed{x}_{3}\Big)+ \frac{B_{1}(D-2\Delta_{1}-1)\slashed{x}_{1}\delta^{IJ}}{(x_{3}^2)^{\Delta_{1}+1/2}(x_{1}^{2})^{a}}
\een
In obtaining the second line above we have used the following results:
\begin{eqnarray}
&&\left<V_{1}^{I}(x_{2})\bar{V}^{J}_{1}(x_{3})\right>=\frac{\slashed{x}_{23}}{(x_{23}^{2})^{\Delta_{1}+1/2}}\,,  \\
&&x_{12}^{\mu}\partial_{2\mu}\left (\frac{\slashed{x}_{23}}{(x_{23}^2)^{\Delta_{1}+1/2}}\right)=\frac{1}{(x_{23}^2)^{\Delta_{1}+3/2}}\left( \slashed{x}_{12}x^{2}_{23}(\frac{1}{2}-\Delta_{1})-(\Delta_{1}+\frac{1}{2})\slashed{x}_{23}\slashed{x}_{12}\slashed{x}_{23}\right)\,\\
&&\slashed{x}_{12}\slashed{\partial}_{2}\left (\frac{\slashed{x}_{23}}{(x_{23}^2)^{\Delta_{1}+1/2}}\right)=\frac{(D-2\Delta_{1}-1)\slashed{x}_{12}}{(x_{23}^2)^{\Delta_{1}+1/2}}\,.
\end{eqnarray}
But from eqn. (\ref{3pt1}),
\ben\label{3pt1series}
\left<V_{2p}(x_{1})V_{2p+1}^{I}(0)\bar{V}_{1}^{J}(x_{3})\right>_{|x_{1}|\ll|x_{3}|}&=& g_{1} \frac{\slashed{x}_{1}\slashed{x}_{3}}{(x_{1}^{2})^{a+1/2}(x_{3}^{2})^{b+1/2+c}}\left[1+\left(c+\frac{1}{2}\right)\frac{\left(\slashed{x}_{1}\slashed{x}_{3}+\slashed{x}_{3}\slashed{x}_{1}\right)}{x_{3}^{2}}+...\right] \nn\\&&+g_{2}\frac{ \slashed{x}_{13}}{(x_{1}^{2})^{a}(x_{3}^{2})^{b+c+1/2}}\left[1+c\frac{\left(\slashed{x}_{1}\slashed{x}_{3}+\slashed{x}_{3}\slashed{x}_{1}\right)}{x_{3}^{2}}+...\right]\,.
\een
Comparing the above equation with eq. (\ref{ope1series}) we get,
\begin{eqnarray}
A_{0}=g_{2}\,,\nonumber\\
B_{0}\left(\frac{1}{2}-\Delta_{1}\right)+B_{1}\left(D-2\Delta_{1}-1\right)= -c g_{2}\,,\\
B_{0}\left(\frac{1}{2}+\Delta_{1}\right)=c g_{2}\,.\nonumber
\end{eqnarray}
Since the tensor structure of the first term in eq. (\ref{3pt1series}) does not have any matching with the tensor structures appearing in eq. (\ref{ope1series}), we can set $g_{1}=0$. 
Finally we have,  
\begin{equation}\label{B1}
B_{0}=\frac{\left(\Delta_{1}+\Delta_{2p}-\Delta_{2p+1}\right)A_{0}}{\left(2\Delta_{1}+1\right)}, \hspace{1.0cm} B_{1}=\frac{B_{0}}{\left(2\Delta_{1}+1-D\right)}\,.
\end{equation}
Next we consider the following 3 pt. function
\begin{equation}\label{3pt2}
\left<V_{2p+1}^{I}(x_{1})V_{2p+2}(x_{2})\bar{V}_{1}^{J}(x_{3})\right>= g'_{1} \frac{\slashed{x}_{12}\slashed{x}_{32}\delta^{IJ}}{(x_{12}^{2})^{a'+1/2}(x_{23}^{2})^{b'+1/2}(x_{31}^{2})^{c'}}+ g'_{2}\frac{\slashed{x}_{13}\delta^{IJ}}{(x_{12}^{2})^{a'}(x_{23}^{2})^{b'}(x_{31}^{2})^{c'+1/2}}\,,
\end{equation}
where, 
\begin{eqnarray}
a'= \frac{\left(\Delta_{2p+1}+\Delta_{2p+2}-\Delta_{1}\right)}{2}\,,\nonumber\\
b'=\frac{\left(\Delta_{2p+2}+\Delta_{1}-\Delta_{2p+1}\right)}{2}\,,\\
c'=\frac{\left(\Delta_{1}+\Delta_{2p+1}-\Delta_{2p+2}\right)}{2}\,. \nonumber
\end{eqnarray}
In this case using the OPE we get 
\begin{equation}
V_{2p+1}^{I}(x_{1})\times V_{2p+2}(0)\supset \left(\frac{\slashed{x}_{12}}{(x_{12}^{2})^{a'}}D(x_{12},\partial_{2})V_{1}^{I}(x_{2})\right)_{x_{2}=0}\,.
\end{equation}
Therefore,
\ben\label{ope2series}
\left<V_{2p+1}^{I}(x_{1})V_{2p+2}(0)\bar{V}_{1}^{J}(x_{3})\right>_{|x_{1}|\ll|x_{3}|} &\sim& \left(\frac{1}{(x_{12}^{2})^{a'}}D(x_{12},\partial_{2})\left<V_{1}^{I}(x_{2})\bar{V}^{J}_{1}(x_{3})\right>\right)_{x_{2}=0}\,,\nn\\ &=&\frac{A'_{0}(-\slashed{x}_{1}\slashed{x}_{3})\delta^{IJ}}{(x_{1}^{2})^{a'}(x_{3}^{2})^{\Delta_{1}+1/2}}+\frac{\slashed{x}_{1}B'_{0}\delta^{IJ}}{(x_{1}^{2})^{a'}(x_{3}^{2})^{\Delta_{1}+1/2}}\Big(\left(\frac{1}{2}-\Delta_{1}\right)x_{3}^{2}\slashed{x}_{1}\nn\\&&-\left(\frac{1}{2}+\Delta_{1}\right)\slashed{x}_{3}\slashed{x}_{1}\slashed{x}_{3}\Big)+ \frac{\delta^{IJ}B'_{1}(D-2\Delta_{1}-1)x_{1}^{2}}{(x_{3}^2)^{\Delta_{1}+1/2}(x_{1}^{2})^{a'}}
\een
In the limit $|x_{1}|\ll|x_{3}|$ eq. (\ref{3pt2}) becomes
\begin{equation*}
\left<V_{2p+1}^{I}(x_{1})V_{2p+2}(0)\bar{V}_{1}^{J}(x_{3})\right>_{|x_{1}|\ll|x_{3}|} = g'_{1} \frac{\slashed{x}_{1}\slashed{x}_{3}}{(x_{1}^{2})^{a'+1/2}(x_{3}^{2})^{b'+1/2+c'}}\left[1+c'\frac{\left(\slashed{x}_{1}\slashed{x}_{3}+\slashed{x}_{3}\slashed{x}_{1}\right)}{x_{3}^{2}}+...\right]
\end{equation*}
\begin{equation}
  + g'_{2}\frac{ \slashed{x}_{13}}{(x_{1}^{2})^{a'}(x_{3}^{2})^{b'+c'+1/2}}\left[1+\left(c'+\frac{1}{2}\right)\frac{\left(\slashed{x}_{1}\slashed{x}_{3}+\slashed{x}_{3}\slashed{x}_{1}\right)}{x_{3}^{2}}+...\right]\,.
\end{equation}
Comparing the above equation with eq. (\ref{ope2series}), we obtain, 
\begin{eqnarray}
  A'_{0}=-g_{1}\,,\nonumber\\
  B'_{0}\left(\frac{1}{2}-\Delta_{1}\right)+B_{1}\left(D-2\Delta_{1}-1\right)= c' g'_{1}\,,\\
   B_{0}\left(\frac{1}{2}+\Delta_{1}\right)=-c' g'_{1}\,.\nonumber
\end{eqnarray}
This gives, 
\begin{equation}\label{B2}
B'_{0}=\frac{\left(\Delta_{1}+\Delta_{2p+1}-\Delta_{2p+2}\right)A'_{0}}{\left(2\Delta_{1}+1\right)}, \hspace{1.0cm} B'_{1}=\frac{B'_{0}}{\left(2\Delta_{1}+1-D\right)}\,.
\end{equation}
Here, similar arguments as above would set $g'_{2}=0$. This again shows that in the 3 pt. function of two primary fermion operators and a primary scalar operator in general one must keep both tensor structures. Which structure contributes in a specific case depends upon the particular primary operators under consideration. When one of the operators involved in the 3 pt. function is a descendant, the allowed structure is of course determined by the correlator of primary operators. 

\subsection{Recursion relations for the leading order anomalous dimensions}
In the $\epsilon\rightarrow 0$ limit, the OPEs of the interacting theory should go over to the free theory OPEs - eqs. (\ref{fope1}, \ref{fope2}) - and the corresponding 3 pt. functions must match as well. This matching gives
\begin{equation}
A_{0}=f_{2p}, \hspace{0.4cm} B_{1}\alpha = f_{2p}\rho_{2p}\,.
\end{equation}
\begin{equation}
\Rightarrow\frac{\left(\Delta_{1}+\Delta_{2p}-\Delta_{2p+1}\right)}{\left(2\Delta_{1}+1\right)\left(2\Delta_{1}+1-D\right)}\alpha=\rho_{2p}\,.
\end{equation}
We use the following relations
\begin{eqnarray}
\Delta_{1}&=&\frac{1+\epsilon}{2}+ \gamma_1\,,\\
\Delta_{2p}&=&2p\left(\frac{1+\epsilon}{2}\right)+\gamma_{2p}\,,\\
\Delta_{2p+1}&=&(2p+1)\left(\frac{1+\epsilon}{2}\right)+\gamma_{2p+1}\,,\\
\alpha&=&2\sigma\left(\frac{\gamma_{1}}{N-1}\right)^{1/2}\,.
\end{eqnarray}
 to get 
\begin{equation}
\frac{\left(\gamma_{2p}-\gamma_{2p+1}\right)}{2\gamma_{1}} \sigma\left(\frac{\gamma_{1}}{N-1}\right)^{1/2}=-\frac{p}{N-1}\,.
\end{equation}
Writing $\gamma_k(\epsilon)=y_{k,1}\epsilon +y_{k,2}\epsilon ^2 +....$ we get,
\begin{equation}
y_{2p+1,1}-y_{2p,1}=2\sigma p \left(\frac{y_{1,2}}{N-1}\right)^{1/2}\,.
\end{equation}
Using $y_{1,2}=\frac{N-1}{4(N-2)^2}$ this gives,
\begin{equation}\label{rec1}
y_{2p+1,1}-y_{2p,1}=\sigma \frac{p}{N-2}\,.
\end{equation}
Similarily we get for the other case,
\begin{equation}
A'_{0}=f_{2p+1}, \hspace{0.4cm} B'_{1}\alpha = f_{2p+1}\rho_{2p+1}\,,
\end{equation}
\begin{equation}
\Rightarrow\frac{\left(\Delta_{1}+\Delta_{2p+1}-\Delta_{2p+2}\right)}{\left(2\Delta_{1}+1\right)\left(2\Delta_{1}+1-D\right)}\alpha=\rho_{2p+1}\,,
\end{equation}
\begin{equation}
\Rightarrow\frac{\left(\gamma_{2p+1}-\gamma_{2p+2}\right)}{2\gamma_{1}} \sigma\left(\frac{\gamma_{1}}{N-1}\right)^{1/2}=\frac{N-p-1}{N-1}\,,
\end{equation}
which gives 
\begin{equation}\label{rec2}
y_{2p+1,1}-y_{2p+2,1}=\sigma \left(\frac{N-p-1}{N-2}\right)\,.
\end{equation}
Solving the recursion relations eqs. (\ref{rec1}) and (\ref{rec2}) (with $\sigma=1$) we get our desired result, 
\begin{equation}\label{final}
y_{2p,1}= -\frac{p(N-p)}{(N-2)}\,,\,\,\,\,\,\,\,\,\,\,\,\,y_{2p+1,1}=-\frac{p(N-p-1)}{(N-2)}\,.
\end{equation}
Thus we have for the scaling dimensions of these composite operators,
\begin{align}
\Delta_{(\bar{\psi}\psi)^{p}}\equiv \Delta_{2p}=p+p\epsilon-\frac{p(N-p)}{(N-2)}\epsilon + O(\epsilon^{2})\,,\\
\Delta_{(\bar{\psi}\psi)^{p}\psi}\equiv \Delta_{2p+1}=(p+\frac{1}{2})+(p+\frac{1}{2})\epsilon-\frac{p(N-p-1)}{(N-2)}\epsilon + O(\epsilon^{2})\,.
\end{align}
Note, in particular, that the classically marginal operator $(\bar{\psi} \psi)^2$ receives corrections to its conformal dimension only at $O(\epsilon^2)$, since for $p=2$ the second  and third terms in the expression for $\Delta_{(\bar{\psi}\psi)^{p}} $cancel. This is analogous to the bosonic case treated in \cite{Rychkov} where the classically marginal operator $(\phi.\phi)^2$ has the same property.

\section{Other scalar primaries }
In this section we will consider a scalar primary which is not a singlet under the symmetry group $U(\tilde N)$ and calculate its anomalous dimension. In the free theory we consider a scalar of the form
\be
\mathcal O^{(IJ)}=\bar\psi^I\psi^J-\frac{\delta^{IJ}}{\tilde N}\bar\psi^K\psi^K\,.
\ee 
In order to calculate the OPE we need the following correlation functions in the free theory:
\ben
&&\left<\psi^K(x)\bar\psi^I\psi^J(0)\bar\psi^L(z)\right>=-\frac{\delta^{KI}\delta^{JL}\slashed x\slashed z}{x^{2}z^{2}}\,,\\
&&\left<\psi^K(x)\bar\psi^I\psi^J(0)\bar\psi^L(\bar\psi^P\psi^P)(z)\right>=-\frac{(\delta^{IK}\delta^{JL}-2\delta^{KL}\delta^{IJ})\Gamma^\mu(x-z)_\mu}{(x-z)^{2}z^{2}}\,.
\een
Therefore for $x\sim 0$, we get
\ben
&&\left<\psi^K(x)\mathcal O^{(IJ)}(0)\bar\psi^L(z)\right>=-\frac{(\delta^{KI}\delta^{JL}-\frac{1}{\tilde N}\delta^{IJ}\delta^{KL})\slashed x\slashed z}{x^2z^2}\,,\nn\\
&&\left<\psi^K(x)\mathcal O^{(IJ)}(0)\bar\psi^L(\bar\psi^P\psi^P)(z)\right>=\frac{(\delta^{KI}\delta^{JL}-\frac{1}{\tilde N}\delta^{IJ}\delta^{KL})\slashed z}{z^4}\,.
\een
Thus we get the following OPE in the free theory,
\ben
&&\psi^K(x)\times \mathcal O^{(IJ)}(0)\supset \frac{(\delta^{KI}\delta^{JL}-\frac{1}{\tilde N}\delta^{IJ}\delta^{KL})}{ x^2}\l\{\slashed x \psi^L(0)+\frac{x^2}{1-N}\psi^L(\bar\psi^P\psi^P)(0)+...\r\}
\een
Now we proceed as before. We assume that there exists an operator $V_{\mathcal O^{(LM)}}(x)$ at the fixed point corresponding to the operator $\mathcal O^{(LM)}(x)$. Based on the symmetries, the 3-point function involving the scalar at the fixed point is given by
\ben
&&\left<V^I_1(x_1)\bar V^K_1(x_2)V_{\mathcal O^{(LM)}}(x_3)\right>=\frac{\tilde f'\slashed x_{13}\slashed x_{23}(\delta^{IL}\delta^{KM}-\frac{1}{\tilde N}\delta^{LM}\delta^{IK})}{(x_{12}^2)^{\Delta_1-\frac{1}{2}\Delta_{(LM)}}(x_{13}^2x_{23}^2)^{\frac{1}{2}\Delta_{(LM)}+\frac{1}{2}}}
\een
The OPE obtained in (\ref{fixedptOPE}) should hold in the case of scalar fields. All the corresponding coefficients are given in (\ref{coeffconfdimension}) except that $\Delta_2$ is replaced by $\Delta_{(LM)}$. Thus in this case,
\be
V^I_1(x_1)\times V_{\mathcal O^{(LM)}}(x_3)\supset \frac{\tilde f\slashed x_{13}(\delta^{IL}\delta^{MK}-\frac{1}{\tilde N}\delta^{ML}\delta^{IK})}{(x_{13}^2)^{\frac{1}{2}\Delta_{(LM)}+\frac{1}{2}}}\tilde E(x_{13},\p_z)V^K_1(z)|_{z=x_3}+..
\ee
with
\be
\tilde E(x_{13},\p_z)=A'+\l(B'_1 x_{13}^\mu+B'_2\slashed{x}_{13}\Gamma^\mu\r)\p_\mu+\l(C'_1 x_{13}^\mu x_{13}^\nu+C'_2x_{13}^\mu\slashed{x}_{13}\Gamma^\nu+C'_3x_{13}^2\Gamma^\mu\Gamma^\nu\r)\p_\mu\p_\nu+.........
\ee
where the relevant coefficients are
\ben\label{coeffconfdimension2}
&&A'=-1 ,\quad B'_1=-\frac{ \l(\Delta_1-\frac{1}{2}\Delta_{(LM)}\r)}{\Delta_1+\frac{1}{2}},\quad B'_2=\frac{B'_1}{2\Delta_1+1-d},
\een 
We proceed as in previous cases. We find that $\tilde f$ should approach $-1$ in the limit $\epsilon\rightarrow 0$. Furthermore the 3-point function with the descendant has the form
\ben
\l<V^I_1(x)V_{\mathcal O^{(LM)}}(0)\bar V^P_3(z)\r>&\sim&\frac{\tilde f\slashed{x}(\delta^{IL}\delta^{KM}-\frac{1}{\tilde N}\delta^{LM}\delta^{IK})}{(x^2)^{\frac{1}{2}\Delta_{(LM)}+\frac{1}{2}}} B_2'(\slashed{x}\,\Gamma^\mu\p_\mu)\l<V^K_1(0)\bar V^P_3(z)\r>\,,\nn\\&&=\frac{\tilde f(\delta^{IL}\delta^{KM}-\frac{1}{\tilde N}\delta^{LM}\delta^{IK}) B_2'\alpha}{(x^2)^{\frac{1}{2}\Delta_{(LM)}-\frac{1}{2}}}\l<V^K_{3}(0)\bar V^P_{3}(z)\r>\,.
\een
Now comparing with the free correlator, we find that
\be
\tilde f B_2'\alpha=-\frac{1}{N-1}\,.
\ee
and 
\be
B_2'=\frac{\pi(\gamma_1-\frac{1}{2}\gamma_{(LM)})}{\gamma_1},\quad \gamma_1\sim y_{1,2}\epsilon^2,\quad \gamma_{(LM)}\sim y_{(LM),1}\epsilon\,.
\ee
which implies that
\be
y_{(LM),1}=\frac{1}{N-2}\,.
\ee
Therefore the leading order anomalous dimension is
\be
\gamma_{\mathcal O^{(IJ)}}=\frac{1}{N-2}\epsilon\,.
\ee
We could not find a check for this result in the literature. It would be interesting to compare this new result against a perturbative computation of the anomalous dimension.

\section{Discussion}
In this note we have computed, to first order in the epsilon expansion, the anomalous dimensions of a class of composite operators in the Gross-Neveu model. As emphasised earlier, we have done the computation without using the usual perturbative techniques. The primary input was conformal symmetry, which fixed for us the two and three point functions and the required OPEs. The main results,  which, to our knowledge, have not been known before, are given in eq. (\ref{final}), . It is to be emphasised that the methods used here can fix only the {\it leading} order anomalous dimensions. It would be interesting to extend the computations to second order in $\epsilon$. As discussed in \cite{Rychkov} for the case of the $O(N)$ bosonic vector model, two and three point functions would not suffice for the higher order computation and one would require conformal bootstrap of the four point functions to extract further information. How conformal symmetry can be used together with OPE associativity to deduce anomalous dimensions at higher orders in $\epsilon$ remains an interesting open problem (see \cite{Aninda,Gupta} for recent work in this direction).

\section*{Acknowledgements}
We thank Aninda Sinha and Kallol Sen for useful discussions and collaboration in the early stages of this work.  RG thanks CHEP at IISc for hospitality during the initial stages of this work.

\begin{appendices}
\section{Computation of $f_{2p}$ and $\rho_{2p}$} 
Here we follow the diagrammatic method \cite{Chetan} to compute the combinatorial factors appearing in the OPEs (\ref{fope1}), (\ref{fope2}). A typical diagram will look like 
\begin{figure}[H]
\begin{center}
\includegraphics[scale=0.34]{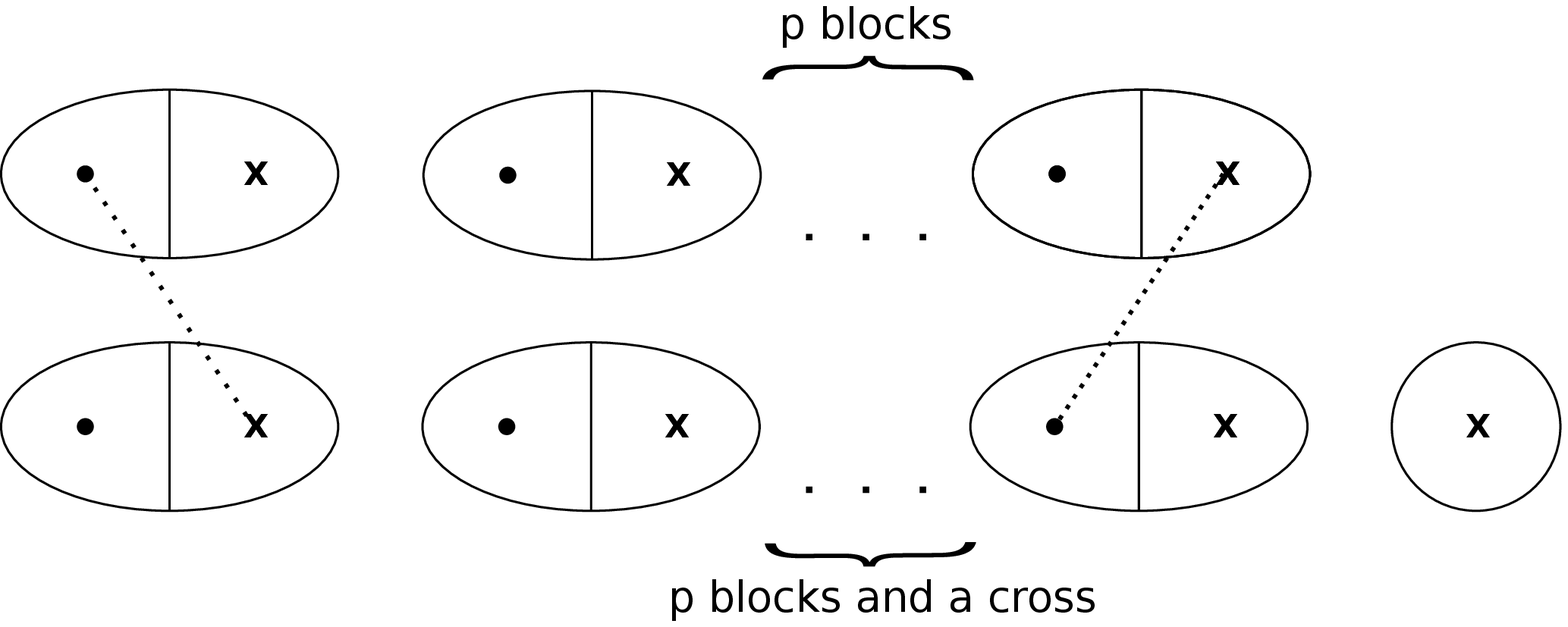}
\end{center}
\end{figure}
where each block refers to a $\overline{\psi}.\psi$ pair, $\overline{\psi}$ is denoted by $\bullet$ and $\psi$ by $\times$, and each line denotes a contraction.
Further we follow the convention that the top row corresponds to the operator at $x$ and the lower one corresponds to the operator at origin. In order to compute the combinatorial factors we need to count all possible contractions, carefully picking up (-1) factors whenever we move a fermionic operator through the other operator. To compute the combinatorial factors our strategy will be to set up recursion relations. For this we contract one block at a time from the top row with the blocks at the bottom row. The contribution of a given contraction is given by following rules:
\begin{enumerate}
\item While contracting we always keep the block at the position $x$ on the left of the block at the origin.
\item {A diagram involving a complete loop will give a factor of $+N$ to the combinatorial factor. 
\begin{figure}[H]
\begin{center}
\includegraphics[scale=0.34]{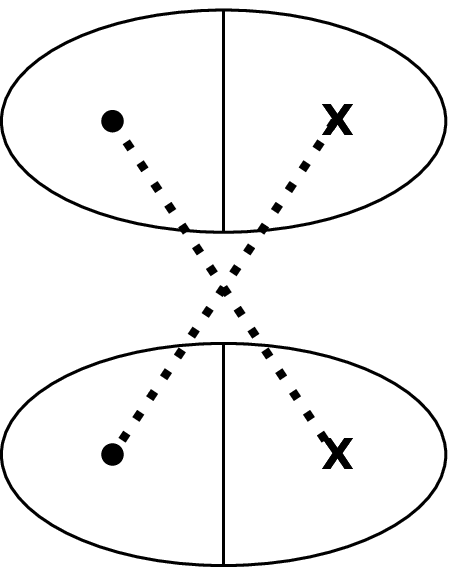}
\end{center}
\end{figure}}
\item {Two blocks contracting to the same block results in one block with $-1$ factor.
\begin{figure}[H]
\begin{center}
\includegraphics[scale=0.34]{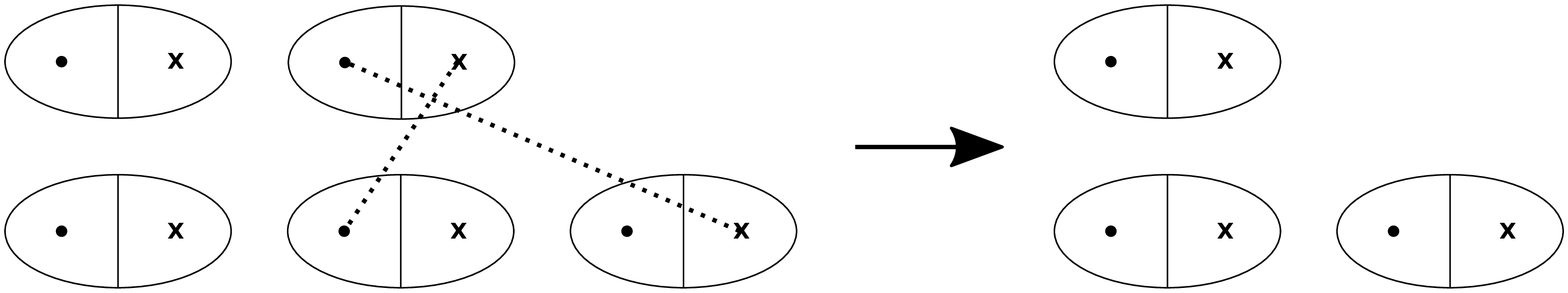}
\end{center}
\end{figure}}
\end{enumerate}
To compute $f_{2p}$ we need to contract $p$ pairs of $\psi$'s and finally leave $\psi^I$. We will get the following three diagrams. In these diagrams the top row has $p$ blocks at $x$ and bottom row has $p$ blocks together with one $\times$ at origin.
\begin{figure}[H]
\begin{center}
\includegraphics[scale=0.34]{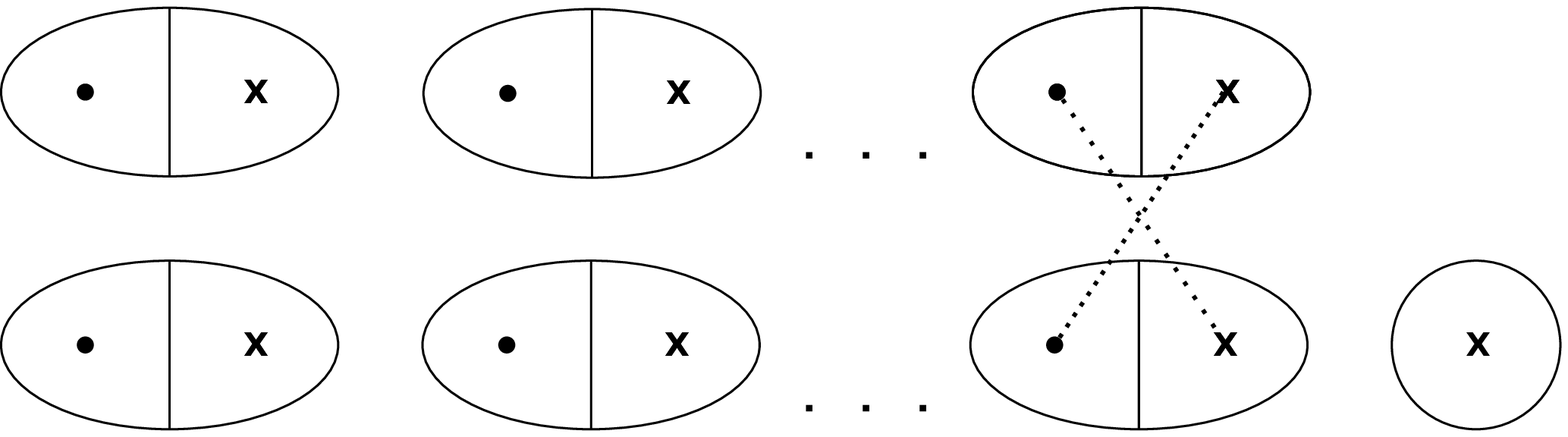}
\end{center}
\end{figure}
The contribution of the above diagram is $p\tilde{N}\Tr{\mathbb{I}} = p N$.
\begin{figure}[H]
\begin{center}
\includegraphics[scale=0.34]{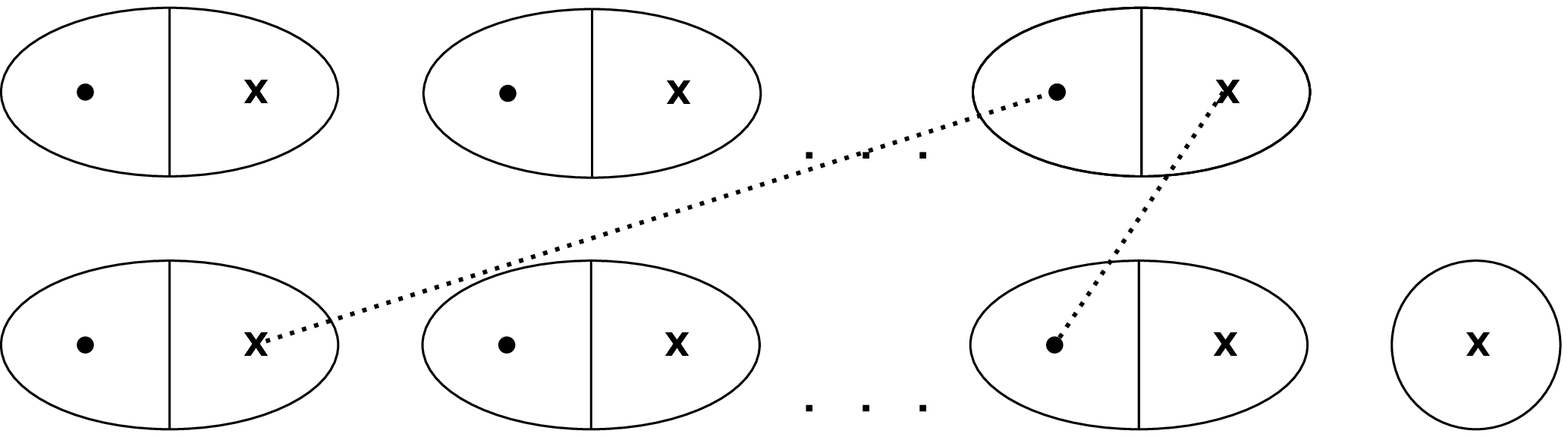}
\end{center}
\end{figure}
The contribution of the above diagram is just $-p(p-1)$,
\begin{figure}[H]
\begin{center}
\includegraphics[scale=0.34]{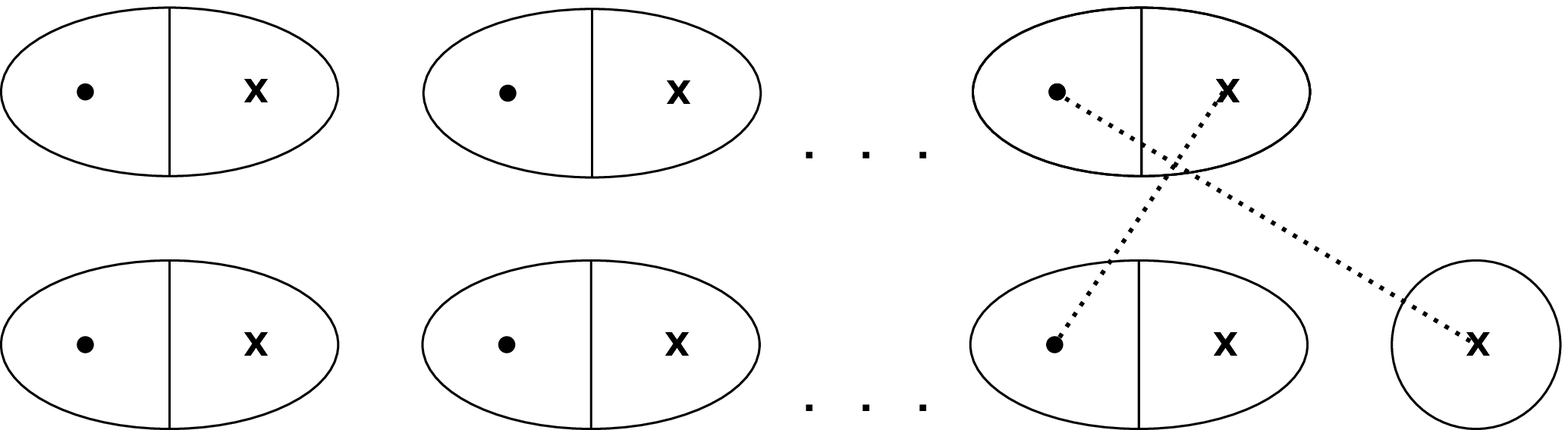}
\end{center}
\end{figure}
The contribution of this diagram is just $-p$. Therefore the recursion relation is 
\be
f_{2p} = [pN - p(p-1) -p] f_{2(p-1)} = [pN - p^2]f_{2(p-1)}
\ee
For $p=1$, we can see that $f_2 = (N-1)$, so 
\be
 f_{2(p)} = \prod_{i=1}^{p}i(N-i)\,.
 \ee
To get a recursion relation for $f_{2p} \rho_{2p}$, the strategy is to first draw one line, and then proceed recursively by drawing two lines. 
To draw one line, we have three diagrams, but the first two diagrams 
\begin{figure}[H]
\includegraphics[scale=0.34]{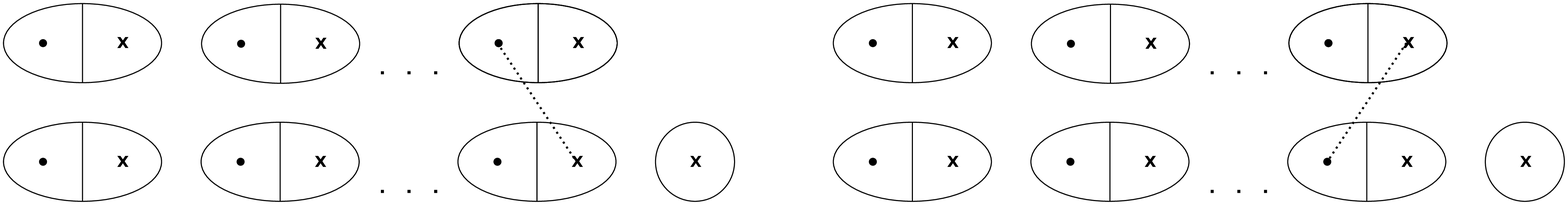}
\end{figure}
cancel each other, and only the third diagram 
\begin{figure}[H]
\includegraphics[scale=0.34]{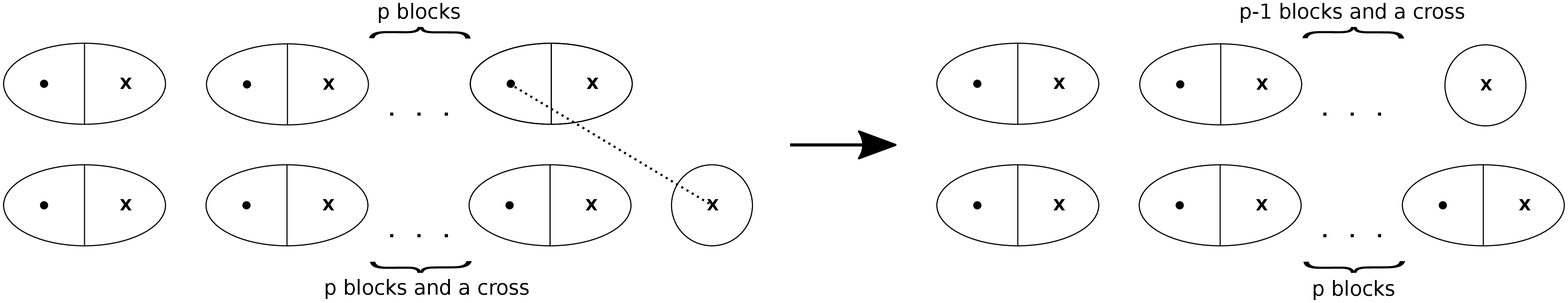}
\end{figure}
will contribute as $-p$. Next we need to compute the contribution, $g_{2p} $, from the new diagram,
\begin{figure}[H]
\begin{center}
\includegraphics[scale=0.34]{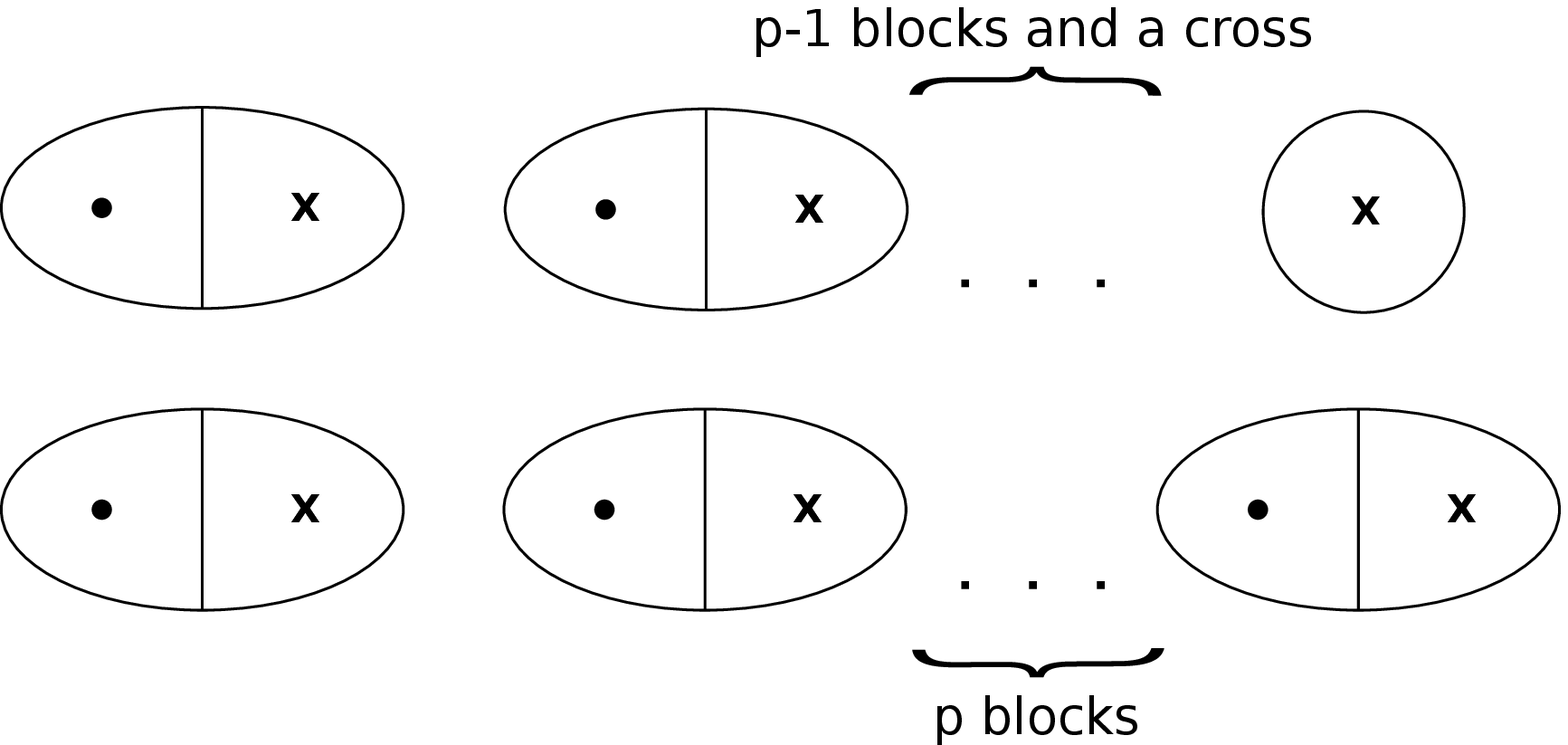}
\end{center}
\end{figure} 
which can be recursively reduced by contracting one block on the top and bottom row, by drawing two lines, 
\begin{figure}[H]
\includegraphics[scale=0.34]{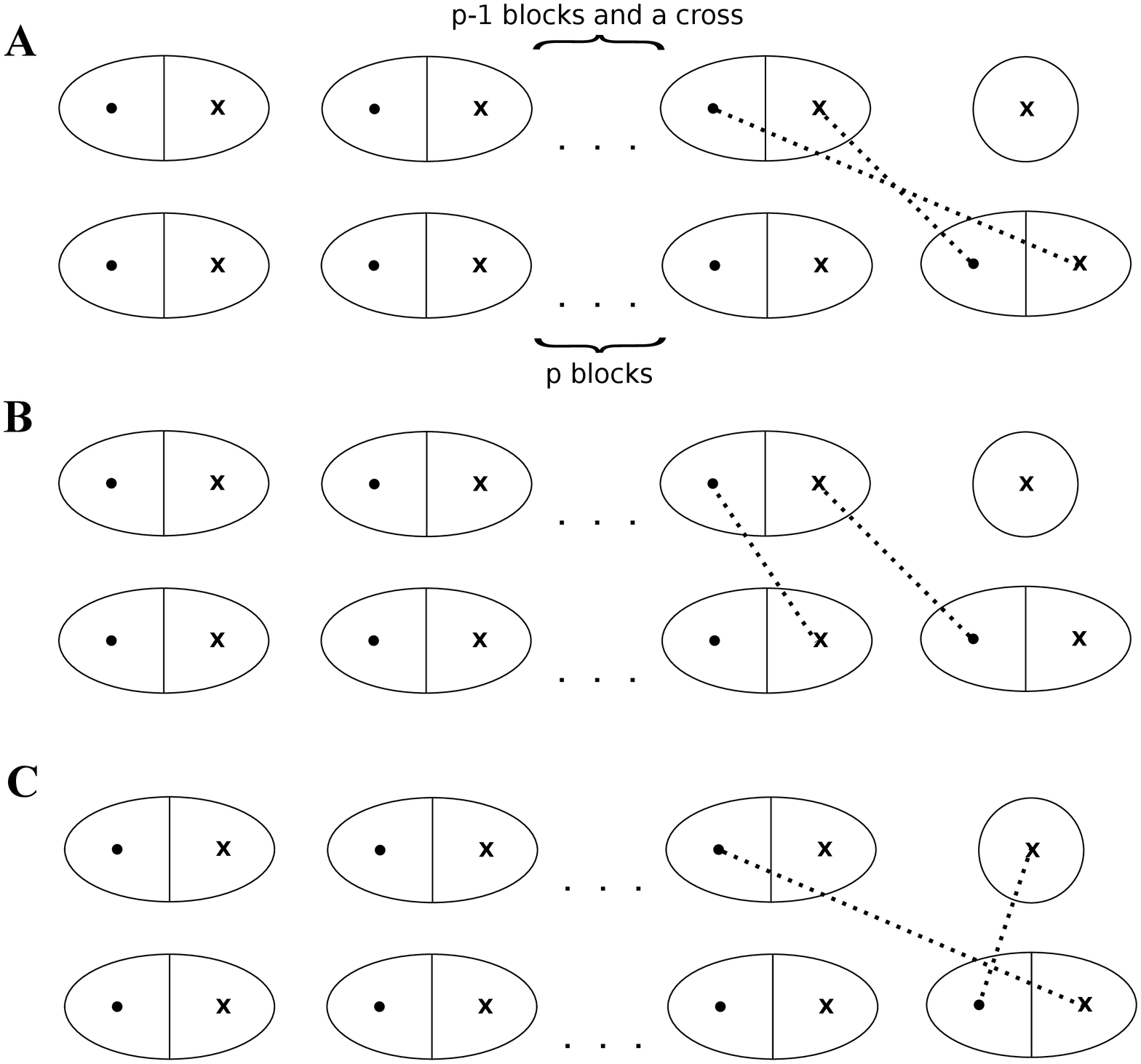}
\end{figure}
The contribution from  $\mathbf{A}$  is $pN$, from $\mathbf{B}$ is $-p(p-1)$ and from $\mathbf{C}$ is $-p$. Therefore the recursion relation for $g_{2p}$ is 
\be
g_{2p} = \l(pN - p^2\r) g_{2(p-1)}\,.
\ee
Knowing $g_{2} = 1$, 
\be 
g_{2p} = \prod_{i=2}^{p} i\l(N-i\r)\,.
\ee
Since, $f_{2p}\rho_{2p} = -p (g_{2p})$, we get
\be
\rho_{2p} = \dfrac{-p (g_{2p})}{f_{2p}} = -\dfrac{p}{N-1}\,.
 \ee
\subsection{Computation of $f_{2p+1}$ and $\rho_{2p+1}$}
The setup is, 
\begin{figure}[H]
\begin{center}
\includegraphics[scale=0.34]{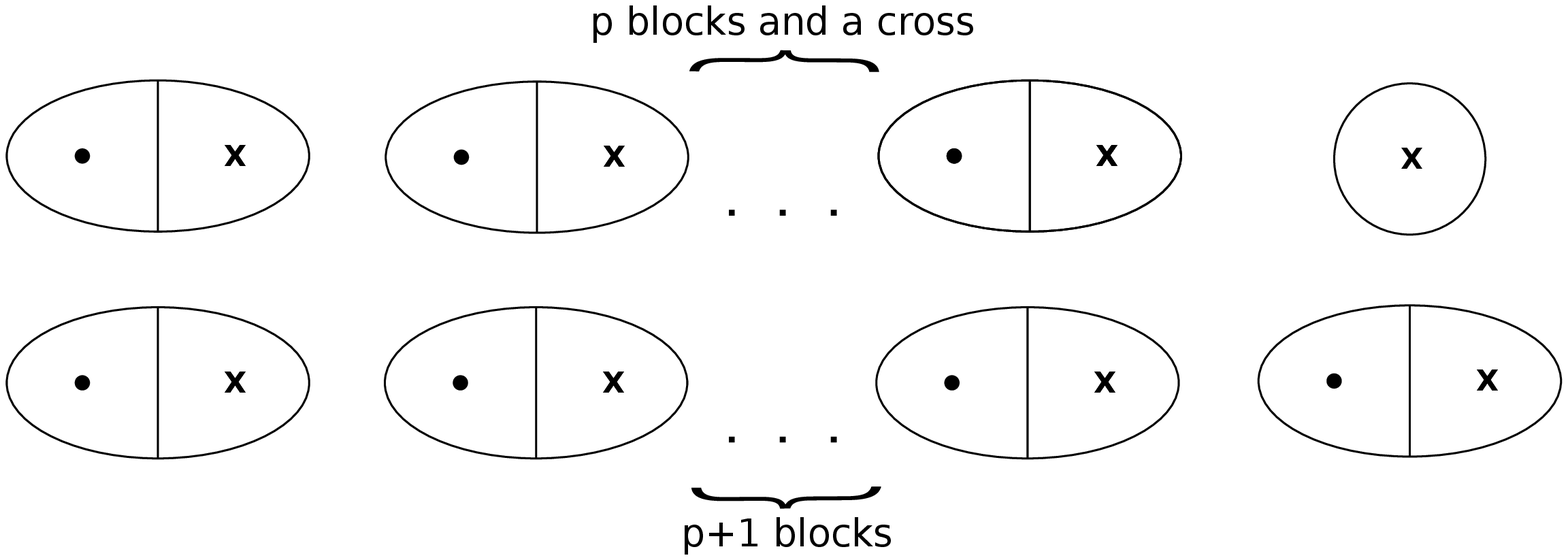}
\end{center}
\end{figure}
and we need to draw $(2p+1)$ lines. Analogous to the $\rho_{2p}$ computation, we first contract one line involving the cross, to give a factor of $+(p+1)$ and the diagram
\begin{figure}[H]
\begin{center}
\includegraphics[scale=0.34]{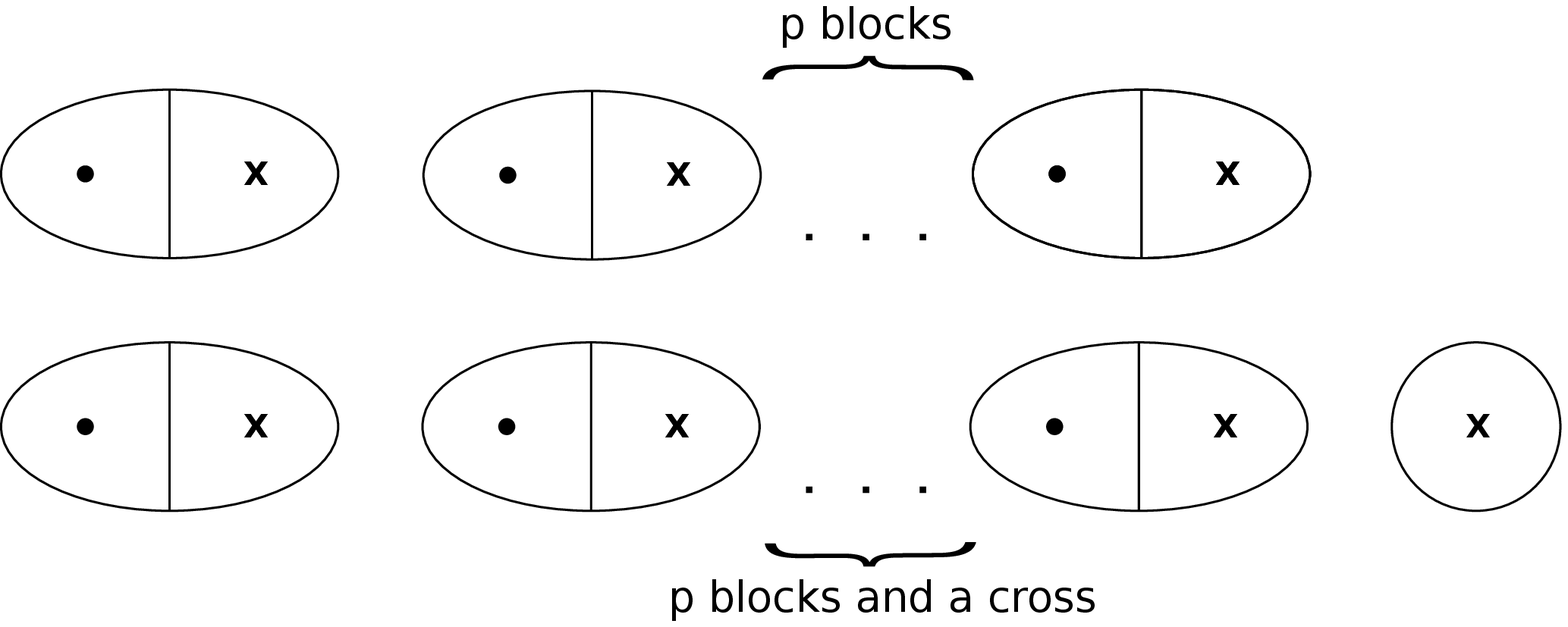}
\end{center}
\end{figure}
which is nothing but $f_{2p}$. Therefore, 
\be
f_{2p+1} = (p+1 )f_{2p} = (p+1)\prod_{i=1}^{p} i(N-i)\,.
\ee
For the computation of $f_{2p+1} \rho_{2p+1} = \tilde g_{2p+1}$, we notice that the diagram is exactly same as that for $g_{2(p+1)}$. Therefore, 
\be
\tilde g_{2p+1} =g_{2(p+1)} =\prod_{i=2}^{p+1} i(N-i)\,.
\ee
So, 
\be
\rho_{2p+1} = \dfrac{\tilde g_{2p+1}}{f_{2p+1}} = \dfrac{N - (p+1)}{N-1}\,.
\ee

\end{appendices}

\end{document}